\documentclass[a4paper,10pt]{article}
\usepackage[utf8x]{inputenc}

\title{Supersymmetric Chern-Simons
 Theory in Presence of a Boundary}
\author{Mir Faizal and Douglas J Smith \\ 
  Department of Mathematical Sciences,\\
Durham University,\\  Durham, DH1 3LE, United Kingdom    
 }

\begin{document}

\maketitle

\begin{abstract}
In this paper we analyse super-Chern-Simons theory in $\mathcal{N} =1$ superspace
formalism, in the presence of a boundary. 
 We  modify the Lagrangian  for the Chern-Simons theory  
in such a way that it is supersymmetric even in the presence of a boundary. 
Also, even though the Chern-Simons theory is not gauge 
 invariant in the presence of a boundary, if it is suitable 
coupled to a gauged Wess-Zumino-Witten model, 
 then the resultant theory can be made gauge invariant.  
Thus, by suitably adding  
 extra boundary degrees of freedom, the gauge and supersymmetry variations
 of the boundary theory  exactly cancel the boundary 
terms generated by the variations of the bulk Chern-Simons theory.
We also discuss how this can be applied to the 
ABJM model in $\mathcal{N} =1$ superspace, and we then describe the
BRST and anti-BRST symmetries of the
resultant gauge invariant supersymmetric theory.

\end{abstract}
\section{Introduction}
The ABJM   theory is  thought to describe the world-volume of multiple M2-branes in 
M-theory at low energies \cite{ABJM, 1}.   It is a
 three dimensional Chern-Simons-matter theory with gauge group
 $U(N)_k \times U(N)_{-k}$ at levels $k$ and $-k$ on 
the world-volume of $N$ M2-branes placed at the fixed point of $R^8/Z_k$. 
Although this construction  explicitly realises only $\mathcal{N} = 6$
 supersymmetry, the supersymmetry is expected to be  
enhanced to full $\mathcal{N} =8$ supersymmetry for $k =1,2$ \cite{su}. 
The ABJM theory coincides with the Bagger-Lambert-Gustavsson (BLG) action
\cite{BL1, BL2, BL3, blG}, based on the Basu-Harvey equation \cite{5},
for the only known example of the Lie $3$-algebra. 

The BLG model has been analysed in the 
 $\mathcal{N} = 1$ superfield formalism  \cite{14}.   
First, an octonionic self-dual
tensor is used to construction a real super-potential with manifest $SO(7)$ invariance. 
Then for specially chosen 
couplings, the component
action coincides with the BLG action,
 and hence the full $SO(8)$ symmetry is restored.
After reduction using the novel Higgs mechanism \cite{14m},
 Higher-derivative corrections to super-Yang-Mills on D2-branes
were analysed in the $\mathcal{N} =1$
 superspace formalism \cite{14abjm}.
Chern-Simons theory with $\mathcal{N} =1$ supersymmetry 
has also been studied in relation to axion gauge symmetry which occurs
in supergravity theories arising from flux compactifications of 
superstrings and Scherk-Schwarz 
generalised dimensional reduction in M-theory \cite{12}. 

The ABJM and BLG actions are formulated for M2-branes without a boundary. However, it
is of interest to allow the inclusion of a boundary. Such boundaries correspond
to M2-branes ending on other objects in M-theory. In \cite{BCIntMem} appropriate
boundary conditions were derived for the ABJM and BLG actions, describing
M2-branes ending on M5-branes, M9-branes or gravitational waves. Boundary
conditions in the presence of background flux were derived in \cite{ChuSehmbi}.
The M5-brane
is of particular interest, and certainly one motivation for studying open M2-branes is to learn about the physics of the
M5-brane.
For example,
by considering a system of M2-branes ending on an M5-brane with a constant $C$-field
turned on, the BLG model was used to motivate a novel quantum geometry on the M5-brane world-volume
\cite{d12}. Another interesting relation between multiple M2-branes and the
M5-brane is the identification of the BLG action (with Nambu-Poisson 3-bracket) as the M5-brane action with a
large worldvolume C-field, as reviewed in \cite{M5BLG}. While these results involve a model for multiple M2-branes,
we note that earlier work using the action for single open M2-branes suggested
a form of non-commutative string theory on the M5-brane worldvolume
\cite{NCS1, NCS2, NCS3}. It would be interesting to understand how these
results arising from different approaches are related.

One of our motivations is to make further progress towards a superspace
description of the ABJM action with a boundary. Rather than specifying boundary
conditions as in \cite{BCIntMem, ChuSehmbi}, the idea here is to add additional
boundary terms and degrees of freedom to make the action consistent. The
prescription is motivated by the symmetries of the bulk action. In particular,
we follow the general prescription given in \cite{21} to add boundary terms so
that half the bulk supersymmetry is preserved. This procedure has been applied
to supersymmetric Abelian Chern-Simons theories in \cite{21} and particularly
to various models including Chern-Simons matter theories and the ABJM model
in \cite{MemBdry}. However, in addition to supersymmetry, it is necessary to
consider preservation of gauge symmetry. This issue was considered, with the aim
of describing
the physics of multiple self-dual strings in \cite{20}. In doing so bosonic Chern-Simons theory on 
a manifold with a boundary was analysed. 
 It was found that even though 
the Chern-Simons theory was not gauge invariant by itself in the presence of a  boundary,
 the sum of it with a Wess-Zumino-Witten model living on the boundary was gauge invariant.
Thus,  new degrees of freedom  were identified on the boundary
 and these degrees of freedom generated a
$U(2N)\times U(2N)$ Kac-Moody current algebra. While it is possible to introduce
the fermionic sector and derive a supersymmetric action in component form, it
seems somewhat natural to derive a manifestly supersymmetric gauge invariant
action, in some sense combining the results of \cite{MemBdry} and \cite{20}.
This will be the result of section~\ref{ABJM}, although for simplicity we
limit ourselves to $\mathcal{N} = 1$ superspace and don't address the issue of
a background $C$-field as there has been limited progress in extending the ABJM
action to include coupling to a general $C$-field \cite{ABJMC1, ABJMC2, ABJMC3}. Because the issue of
preservation of gauge symmetry is specific to the Chern-Simons term, this is
considered
separately in section~\ref{CS}.

While there is a well-known connection between $(2+1)$-dimensional (topological)
Chern-Simons
theories and $(1+1)$-dimensional CFTs \cite{QFTJonesPoly}, the situation is
less clear for Chern-Simons matter theories. As shown in \cite{MooreSeiberg, EMSS}
for pure Chern-Simons theory with suitable boundary conditions, a component of
the gauge field, say $A_0$, appears linearly in the action and so can be
integrated out, imposing the constraint $F_{12} = 0$. This constraint can be
solved explicitly (e.g. for a manifold of the form of a disk for each constant time slice) and the result is a $(1+1)$-dimensional WZW model where the bulk
gauge potential has been replaced by the boundary gauge degrees of freedom.
Now Chern-Simons matter theories are not topological so we should not expect
such a connection to $(1+1)$-dimensional CFTs. Of course, in cases such as ABJM
theory where the Chern-Simons matter theory is conformal, the boundary theory
may still be conformal. However, an important difference to the pure
Chern-Simons case is that due to the gauged scalar kinetic terms, $A_0$ will
no longer appear as a Lagrange multiplier -- even the classical equation of
motion will couple $F_{12}$ to the scalars rather than simply requiring
$F_{12} = 0$. We therefore cannot expect the Chern-Simons action to be replaced
by a WZW model in general. However, it is possible to use the principle of
gauge invariance in the presence of a boundary to couple the Chern-Simons
theory to a boundary theory. The general result is a gauge invariant action coupling the
Chern-Simons gauge potential to a boundary WZW model, and which reproduces the
pure WZW action when starting from a pure Chern-Simon action \cite{20}.

Supersymmetric
Chern-Simons theories have also been studied as interesting examples 
of the $AdS_4/CFT_3$ correspondence \cite{6,7,8,9,10}.  
Three dimensional $\mathcal{N} = 1$ super-conformal field theories have the property of being  
 supersymmetric without having  any holomorphic
property. This is a peculiarity of the $AdS_4/CFT_3$ correspondence with respect to the
usual $AdS_5/CFT_4$. Thus, the results of this paper may be useful in analysing certain aspects 
of the  $AdS_4/CFT_3$ correspondence.   

We  need to fix a gauge before we can quantize 
any  theory which has a gauge symmetry
associated with it. This is done 
 by the addition of  
a gauge fixing term and a ghost term to the
original action.  The action thus obtained is invariant
under two new symmetries called the BRST symmetry \cite{brst1, brst12}
 and the anti-BRST symmetry \cite{antibrst1}.  
These symmetries are important to show the unitarity 
of the $\mathcal{S}$-matrix and thus  the 
consistency of the theory at quantum level \cite{qbrst1}. 
The BRST symmetry of the bosonic Chern-Simons theory has been thoroughly
 investigated
\cite{16,17} and the BRST symmetry of the $\mathcal{N} = 1$   
 Chern-Simons theory   has been analyzed in the superspace 
formalism \cite{18, 19}.
The BRST and the anti-BRST symmetries of the ABJM theory 
have also been  studied
\cite{mf012}.
In this paper we will  analyse the BRST and the anti-BRST symmetries 
of the ABJM theory in presence of a boundary.

\section{Properties of Super-Covariant Derivatives}

In this section we shall first review the properties of the super-covariant derivatives for 
 non-Abelian $\mathcal{N} = 1$ gauge fields in three dimensions \cite{1001SUSY}. Then we shall 
analyse the effect of having a boundary by
 generalising the results of \cite{21} to non-Abelian case. In order to
analyse the properties of the super-covariant 
derivatives, we first introduce $\theta_a$ as
two component anti-commuting parameters with odd Grassmann parity and let 
\begin{equation}
 \theta^2 = \frac{1}{2} \theta_a C^{ab} \theta_b =  \frac{1}{2} \theta^a \theta_a.
\end{equation}
The anti-symmetric tensors $C^{ab}$ and $C_{ab}$  
can be used to raise and lower spinor indices, and they satisfy 
 $C_{ab}C^{bc} = \delta^c_a$.
Now if $T_A$ are Hermitian generators of a Lie algebra 
$[T_A, T_B] = i f_{AB}^C T_C$, 
in the adjoint representation, 
then matter fields can be represented by 
matrix valued  complex scalar superfields $X$ and $X^{\dagger}$ suitably
 contracted with the generators of this Lie algebra, 
 $X= X^A T_A,$ and $ 
X^{\dagger}= X^{\dagger A} T_A$.
Let these superfields transform under infinitesimal gauge
 transformations as
\begin{eqnarray}
  \delta X&=&  i\Lambda X,\nonumber\\
\delta X^{\dagger} &=& -i X^{\dagger}\Lambda .
\end{eqnarray}
where $\Lambda = \Lambda^A T_A$ and the product 
of these fields is actually a commutator.
 Now the  super-derivative,  given by 
\begin{equation}
 D_a = \partial_a + (\gamma^\mu \partial_\mu)^b_a \theta_b,
\end{equation}
of these superfields does not transform 
 like the original superfields. But 
 we can define a super-covariant derivative for  these superfields 
by requiring it   to transform like the original superfields. 
 Thus, we obtain the following expression for the     
 super-covariant derivative of these superfields
\begin{eqnarray}
  \nabla_a  X&=& D_a X -i \Gamma_a X,\nonumber\\
\nabla_a X^{\dagger} &=& D_a X^{\dagger} + iX^{\dagger} \Gamma_a , 
\end{eqnarray}
 where $\Gamma_a$ is a matrix valued spinor
 superfield suitable contracted with generators of a Lie algebra,  $ \Gamma_a = \Gamma_a^A T_A $. 
If this matrix valued spinor superfield is made to transform under 
gauge transformations as 
\begin{equation}
 \delta \Gamma_a = \nabla_a \Lambda. \label{sgt}
\end{equation}
then the  super-covariant derivative of the
 scalar superfields $X$ and $X^{\dagger}$ indeed transforms
 under gauge   transformations like the original fields,
\begin{eqnarray}
  \delta \nabla_a X&=&  i\Lambda  \nabla_a X ,\nonumber\\
\delta \nabla_aX^{\dagger} &=& -i  \nabla_aX^{\dagger}\Lambda.
\end{eqnarray}

Now we can derive certain properties of these super-covariant derivatives.
 The Abelian version of these properties is given in \cite{21}.
Now define the components of this superfield  $\Gamma_a$ to be
\begin{eqnarray}
 \chi_a = [\Gamma_a]_|, && A =  -\frac{1}{2}[\nabla^a \Gamma_a]_|, \nonumber \\ 
A^\mu = - \frac{1 }{2} [ \nabla^a (\gamma^{\mu })_a^b \Gamma_b ]_|, &&
 E_a =\frac{1 }{2}[\nabla^b \nabla_a  \Gamma_b]_|, \label{csf}
\end{eqnarray}
where $'|'$  means that the quantity is evaluated at $\theta_a =0$,
and let $\mathcal{D}_\mu$ be the conventional covariant derivative given by 
\begin{equation}
 \mathcal{D}_\mu = \partial_\mu -i A_\mu.
\end{equation}
Then it can be shown by direct computation that the super-covariant derivative  satisfies 
\begin{equation}
 \{\nabla_a , \nabla_b\} = - 2 \nabla_{ab},
\end{equation}
where 
\begin{eqnarray}
  \nabla_{ab} &=& \partial_{ab} - i\Gamma_{ab}, \nonumber \\ 
  \Gamma_{ab} &=& -\frac{i}{2} \left[ D_{(a}\Gamma_{b)} 
- i\{\Gamma_a, \Gamma_b\} \right],
\end{eqnarray}
and  $ \partial _{ab} = (\gamma^{\mu} \partial_\mu)_{ab}$. 
Now as we are studying $\mathcal{N} =1$ superfields 
 in three dimensions the indices $'a'$ are two-dimensional and
 so $[\nabla_a , \nabla_b]$ must be proportional
 to the anti-symmetric tensor $C_{ab}$. Thus, we find 
\begin{eqnarray}
  \nabla_a \nabla_b &=&  \frac{1}{2} \{ \nabla_a, \nabla_b\} + \frac{1}{2}
 [\nabla_a, \nabla_b]  \nonumber \\ &=&  \gamma^\mu _{ab} \mathcal{D}_\mu - C_{ab} \nabla^2.
\end{eqnarray}
The complete antisymmetrisation of three two-dimensional indices vanishes and so we have 
\begin{equation}
 \nabla_a \nabla_b \nabla_c = \frac{1}{2} \nabla_a \{\nabla_b, \nabla_c\} -  
 \frac{1}{2} \nabla_b \{\nabla_a, \nabla_c \} + \frac{1}{2} \nabla_c \{\nabla_a ,\nabla_b \}. 
\end{equation}
Thus, we get 
\begin{eqnarray}
\nabla^a \nabla_b \nabla_a &=& 0, \label{ar} \label{nab}\\ 
\nabla^2 \nabla_a &=&   (\gamma^\mu \nabla )_{a} \mathcal{D}_\mu, \label{q} 
\end{eqnarray}

If we put a boundary at fixed $x^3$,
then $\mu$ splits into $\mu =(m, 3 )$.
The induced value of the super-derivative $D_a$ and the 
super-covariant derivative $\nabla_a$ on the boundary is denoted by 
$D_a'$ and $\nabla_a'$ respectively.  
This boundary super-derivative $D_a'$ is obtained by neglecting 
$\gamma^3 \partial_3$ contributions in $D_a$, 
 \begin{equation}
 D_a' = \partial_a + (\gamma^m \partial_m)^b_a \theta_b.
\end{equation}
The boundary super-covariant derivative $\nabla_a'$ can thus be written as  
\begin{eqnarray}
  \nabla_a'  X'&=& D_a' X' -i \Gamma_a'X',\nonumber\\
\nabla_a' {X^{\dagger}}' &=& D_a'{X^{\dagger}}'
 + i {X^{\dagger}}'\Gamma_a', 
\end{eqnarray}
where $X', {X^{\dagger}}'$ and $\Gamma_a'$ are the induced values of 
the bulk fields $X, X^{\dagger}$ and $\Gamma_a$ on the 
boundary. Any boundary field along with the 
induced value of any quantity e.g.,  $\Lambda$ on the boundary will be denoted by $\Lambda'$.
This  convention will be followed even for component fields of superfields.  
The  matrix valued spinor superfield $\Gamma_a'$ transforms under gauge transformations as follows:
\begin{equation}
 \delta \Gamma'_a = \nabla'_a \Lambda',
\end{equation}
where $\Lambda'$ is the induced value of $\Lambda$ on the boundary. 

Now we define projection operators $P_{\pm}$ as:
\begin{equation}
 (P_{\pm  })^b_a  = \frac{1}{2} (\delta_a^b \pm (\gamma^{3}) ^b_a). \label{a}
\end{equation}
These projection operators can be used to project the super-covariant derivative $\nabla_a$ as, 
\begin{equation}
 \nabla_{ \pm b }= (P_{\pm})^a_b \nabla_a, 
\end{equation}
and  $\nabla'_{\pm b}$ as, 
\begin{equation}
 \nabla'_{\pm b } = (P_{\pm })^a_b \nabla'_a, 
\end{equation}
where $\nabla_{\pm a}'$ is the induced value of $\nabla_{\pm a}$ on the boundary. 
These projected values of the super-covariant derivative can now be shown to satisfy
\begin{eqnarray}
 \nabla_{+a} \nabla_{+b} &=& - (P_+ \gamma^m)_{ab} \mathcal{D}_m, \\
 \nabla_{-a} \nabla_{-b} &=& - (P_- \gamma^m)_{ab} \mathcal{D}_m, \\
 \nabla_{-a} \nabla_{+b} &=&  -(P_-)_{ab} (\mathcal{D}_3 + \nabla^2),\\  
 \nabla_{+a} \nabla_{-b} &=&  (P_+)_{ab} (\mathcal{D}_3 -\nabla^2 ). \label{2}
 \end{eqnarray}
From these relations we can obtain the following algebra for these projected operators. 
\begin{eqnarray}
 \{ \nabla_{ + a}, \nabla_{ +b} \} &=& - 2  (P_+ \gamma^m)_{ab} \mathcal{D}_m, \\
  \{ \nabla_{-a }, \nabla_{-b} \} &=& - 2  (P_- \gamma^m)_{ab} \mathcal{D}_m, \\
 \{ \nabla_{- a}, \nabla_{ +b} \} &=&  -2  (P_-)_{ab} \mathcal{D}_3. \label{algebra}
\end{eqnarray}
It will be useful to write  Eq. $(\ref{2})$ as 
\begin{eqnarray}
 - \nabla_+ \nabla_- &=&  -C^{ab}\nabla_{+a} \nabla_{-b}  \nonumber \\
 &=& - C^{ab}(P_+)_{ab} (\mathcal{D}_3 -\nabla^2 ) \nonumber \\
 & = & -(P_+)^a_a(\mathcal{D}_3 -\nabla^2) \nonumber \\ &=&  (\mathcal{D}_3 -\nabla^2).
\end{eqnarray}
Note that is is also easy to see that the boundary super-derivatives satisfy
similar relations, and that the supersymmetry splits into left- and right-moving
sectors on the boundary since e.g.\
\begin{equation}
(P_{\pm} \gamma^m)_{ab} \mathcal{D}_m = (\gamma^{\pm})_{ab} \mathcal{D}_{\pm}
\end{equation}
where $\gamma^{\pm} = \gamma^0 \pm \gamma^1$ and
$\mathcal{D}_{\pm} = \frac{1}{2} (\mathcal{D}_0 \pm \mathcal{D}_1)$.

We have now reviewed properties of super-covariant derivatives and extended
results in \cite{21} to non-Abelian theories. In the next section we will 
use these results to analyse non-Abelian Chern-Simons theory in the presence of 
a boundary. 

\section{$\mathcal{N}=1$ Chern-Simons  Theory} 
\label{CS}

Before we consider a boundary we will review
$\mathcal{N}=1$ non-Abelian Chern-Simons theory on a manifold without a boundary.
Now the  Lagrangian  for $\mathcal{N}=1$ non-Abelian Chern-Simons theory in superspace formalism 
 can be written (with implicit trace) as \cite{1001SUSY}
\begin{equation}
 \mathcal{L}_{CS,  k} (\Gamma) = -\frac{k}{4\pi}\nabla^2   [\Gamma^a  \Omega_a ]_|.
\label{LCS}
\end{equation}
 where \cite{1001SUSY}
\begin{eqnarray}
 \Omega_a & = & \omega_a - \frac{1}{6}[\Gamma^b, \Gamma_{ab}] \\
 \omega_a & = & \frac{1}{2} D^b D_a \Gamma_b - \frac{i}{2}  [\Gamma^b , D_b \Gamma_a] -
 \frac{1}{6} [ \Gamma^b ,
\{ \Gamma_b , \Gamma_a\} ], \label{omega} \\
 \Gamma_{ab} & = & -\frac{i}{2} \left[ D_{(a}\Gamma_{b)} 
- i\{\Gamma_a, \Gamma_b\} \right] .
\end{eqnarray} 
In Eqs.~(\ref{LCS}) a trace over the 
generators of the Lie algebra   is implied.  
The covariant divergence of $\omega_a$ vanishes~\cite{14abjm}
\begin{equation}
\nabla^a \omega_a =  0.
\label{DivOmega}
\end{equation}
The components of the superfield  $\omega_a$ can now be calculated from 
Eqs. $(\ref{csf})$  and~$(\ref{omega})$,  
\begin{eqnarray}
 [ \nabla^a (\gamma^{\mu })_a^b \omega_b ]_|=  \epsilon^{\mu \nu \rho} F_{\nu \rho},  &&
[\nabla^a \omega_a]_| =0,
 \\ 
 - [\nabla^b \nabla_a  \omega_b]_| = 2 (\gamma^\mu \mathcal{D}_\mu)_a^b E_b, &&[\omega_a]_| =  E_a , \nonumber
\end{eqnarray}
where $\epsilon_{\mu \nu \rho} $ is an anti-symmetric tensor. So  
 the component form for the Lagrangian for $\mathcal{N}=1$
 non-Abelian Chern-Simons 
theory can be written as
\begin{equation}
 \mathcal{L}_{CS,  k} = \frac{k}{4\pi}\left[ \epsilon^{\mu \nu \rho} \left( A_\mu \partial_\nu  A_\rho 
 + \frac{2i}{3} A_\mu A_\nu  A_\rho  \right) + 
 E^a E_a + \mathcal{D}_\mu ( \chi^a(\gamma^\mu)_a^b E_b)\right].
\end{equation}
Now if the full finite gauge transformation  of the superfield $\Gamma_a$ is written as    
\begin{equation}
   \Gamma_a \rightarrow i u \, \nabla_a u^{-1},
\end{equation}
where 
\begin{equation}
 u = \exp ( i \Lambda^A T_A), 
\end{equation}
then the gauge transformation of the superfield $ \omega_a$ will be given by  
\begin{equation}
  \omega_a \rightarrow u \, \omega_a u^{-1}. 
\end{equation}
Under infinitesimal gauge transformations the Lagrangian
for the $\mathcal{N}=1$ non-Abelian Chern-Simons theory
transforms as  
\begin{equation}
  \delta \mathcal{L}_{CS,  k}(\Gamma) =
  - \frac{k}{4\pi} \nabla^2 [(\nabla^a  \Lambda )\omega_a ]_|.
\end{equation}
Now using Eq. (\ref{DivOmega}), we get 
\begin{eqnarray}
 \delta \mathcal{L}_{CS,  k}(\Gamma) & = &
  - \frac{k}{4\pi} \nabla^2 \nabla^a [ \Lambda \omega_a ]_| \nonumber \\
 & = & - \frac{k}{4\pi} (\gamma^\mu  \mathcal{D}_\mu \nabla)^a[ \Lambda \omega_a ]_|.
\end{eqnarray} 
As this is a total derivative, on a manifold without a boundary we have 
\begin{equation}
 \delta \mathcal{L}_{CS,  k} =0. 
\end{equation}
Thus, the $\mathcal{N}=1$ non-Abelian Chern-Simons theory is invariant under these gauge 
  transformations on a manifold without a boundary. 

After reviewing the gauge invariance of the $\mathcal{N}=1$  
non-Abelian Chern-Simons theory on a manifold without a boundary,
 we can now 
discuss the effect of a boundary on it. The effect 
of a boundary in three dimensions 
on the SUSY of a $\mathcal{N}=1$ theories, and in particular how SUSY can be
preserved by adding additional boundary terms 
has been recently studied in
\cite{21}.
The SUSY variation of the Lagrangian for $\mathcal{N}=1$
 non-Abelian Chern-Simons theory transforms into a total derivative, so  
in the absence of a boundary this variation vanishes and the theory is SUSY. 
However, in the presence of a boundary it reduces to 
a boundary term.   This  theory can still be made SUSY by adding a 
boundary term whose SUSY variation cancels the SUSY variation 
of the original action.
The analysis performed for Abelian Chern-Simons theories in
\cite{21, MemBdry}
can be easily generalised to the non-Abelian case for $\mathcal{N}=1$
 SUSY,
with the result that the boundary term whose addition 
will make $\mathcal{N}
=1$ 
non-Abelian Chern-Simons theory SUSY can be written as   
\begin{equation}
\mathcal{L}_{bCS,  k}(\Gamma) =\frac{k}{4\pi} \mathcal{D}_3 
[\Gamma^a \Omega_a ]_|. 
\end{equation}
In component form this term can be written as
\begin{equation}
\mathcal{L}_{bCS,  k} = \frac{k}{4\pi} \mathcal{D}_3\left[ \chi^a E_a  + \frac{i}{6} \chi^a\left[(\gamma^\mu A_{\mu})_{a}^b ,\chi_b \right] \right].
\end{equation}

The SUSY variation of this boundary term exactly 
cancels the SUSY variation of the bulk Lagrangian, so 
 the sum of the bulk Lagrangian and this boundary term  is SUSY, 
\begin{eqnarray}
   \mathcal{L}_{sCS,   k}(\Gamma) &=& \mathcal{L}_{CS,  k} 
+ \mathcal{L}_{bCS,  k} \nonumber \\ &=& 
   \frac{k}{4\pi} (-\nabla^2 + \mathcal{D}_3) [\Gamma^a \Omega_a]_|. \label{t}
\end{eqnarray}
It may be noted that only half of the SUSY of the original theory
 is preserved on the boundary. In this paper we will keep the 
SUSY corresponding to $\nabla_-$ and break the SUSY corresponding to
$\nabla_+$ on the boundary. 

This SUSY Lagrangian with a boundary term is not gauge invariant 
 because following what we did for 
the $\mathcal{N}=1$ non-Abelian Chern-Simons theory on
 a manifold without boundary, 
  the infinitesimal gauge transformation  of this Lagrangian is given by  
\begin{eqnarray}
 \delta \mathcal{L}_{sCS,   k} (\Gamma)
  &=&\frac{k}{4\pi} ( \mathcal{D}_3 - \nabla^2 )
 \nabla^a [ \Lambda \omega_a]_|. 
\end{eqnarray}
Now using Eq. $(\ref{q})$, this   can be written as
\begin{equation}
  \delta \mathcal{L}_{sCS,   k} (\Gamma) = 
\frac{k}{4\pi} ( \mathcal{D}_3\nabla^a -(\gamma^\mu  \mathcal{D}_\mu \nabla)^a
 ) [ \Lambda \omega_a  ]_|.
\end{equation}
As there is a boundary in the $x^3$ direction, we get  
\begin{eqnarray}
\delta \mathcal{L}_{sCS,   k}(\Gamma)  &=& 
 \frac{k}{4\pi} (\mathcal{D}_3\nabla^a -(\gamma^\mu  \mathcal{D}_\mu \nabla)^a) [ \Lambda \omega_a  ]_|
 \nonumber \\ &\sim
& \frac{k}{4\pi}(\mathcal{D}_3\nabla^a - (\gamma^3  \mathcal{D}_3 \nabla)^a)[ \Lambda \omega_a  ]_|,
\end{eqnarray}
where $\sim$ indicates 
that we have neglected the total derivative contribution along  directions other than $x^3$, as they will not contribute.  
Thus, the gauge   transformation of this SUSY Lagrangian  gives a boundary term,   
\begin{eqnarray}
  \delta \mathcal{L}'_{sCS,   k}(\Gamma')  
 &=&\frac{k}{4\pi}( \delta_b^a - (\gamma^3)_b^a  ){\nabla'^b} [ \Lambda' \omega'_a  ]_| \nonumber \\ 
  &=&   \frac{k}{2\pi}(P_- {\nabla}')^a  [ \Lambda' \omega'_a  ]_|. \label{pp}
\end{eqnarray}
This boundary term can be written in component form as 
\begin{equation}
 \delta  \mathcal{L}'_{sCS,   k}  =
 \frac{k}{2\pi}\left(\epsilon^{\mu \nu}   \lambda' F'_{\mu \nu}  
+  ( {\lambda^a}'(\gamma^3)_a^b {E_b}') 
+  (  {\lambda^a}' {E_a}' )\right),
\end{equation}
where $
 \lambda = [\Lambda]_|,\, \,  \lambda_a = [\nabla_a \Lambda]_|,$
and the `prime' notation $\lambda',\lambda_a',A'_{\mu} $ etc.\ denotes the induced values of these fields on the boundary. 
Due to the presence of this boundary term, the 
  $\mathcal{N}=1$ non-Abelian Chern-Simons theory  is not gauge invariant in the presence of a boundary. 

However, it is possible to couple this theory to another boundary theory, 
such that the total Lagrangian, which is 
given by the sum of the Lagrangians of both these theories, is gauge invariant.
 To do so we consider a boundary theory with the following potential term 
\begin{eqnarray}
  \mathcal{L}_{pb, k}(v', \Gamma') = \mathcal{L}_{sCS,   k}(\Gamma^v)
 - \mathcal{L}_{sCS,   k}(\Gamma),
\end{eqnarray}
where $v'$ is a boundary scalar superfield, $v$ is an extension of $v'$
into the bulk and $\Gamma^v$ denotes the gauge transformation of $\Gamma$ by
$v$. For $v$ close to the identity, this is a genuine boundary term, while in
general we can still consider this to only depend on the boundary in the sense
that in the absence of a boundary this term will have no effect since the
normalisation of the Chern-Simons action is chosen so that the
path integral is also invariant under large
gauge transformations.
 See \cite{20} for a more detailed discussion of the bosonic theory.
Now the total Lagrangian
$\mathcal{L}_{sCS,   k}(\Gamma) + \mathcal{L}_{pb, k}(v', \Gamma')$
will clearly be
gauge invariant if $\Gamma^v$ is. This is possible if we require $v$ to
transform under  
gauge transformations as
\begin{equation}
  v \rightarrow vu^{-1}.
\end{equation}

To better understand this boundary Lagrangian, we can consider the case 
where
$\Gamma^a =0$ so that there is no coupling to the bulk fields. In this case
the boundary term $\mathcal{L}_{sCS, k}(\Gamma^a =-i(\nabla^a v)v^{-1})$
gives the
potential term of the ${\cal N} = (1,0)$ WZW model \cite{WZW1, WZW2}
\begin{equation}
\mathcal{L}_{pb, k}(v', \Gamma') =
-\frac{k}{2\pi} (P_- {\nabla}')^a \left[ [ (v^{-1} \mathcal{D}_{+} v), 
(v^{-1} \mathcal{D}_{3} v)]
(v^{-1} \nabla_{-a} v) \right]_|.
\end{equation}
We can now add the following supersymmetric gauge invariant kinetic term for
the boundary sclar superfield $\hat{v} = v'(\theta_{+} = 0)$,
\begin{eqnarray}
  \mathcal{L}_{kb, k}(v', \Gamma') &=&  - \frac{|k|}{2\pi} 
(P_- {\nabla}')^a [ 
(\hat{v}^{-1} \nabla'_{-a} \hat{v}) (\hat{v}^{-1} \mathcal{D}_{+} \hat{v}) ]_|,
\end{eqnarray}
which is a gauging of
the kinetic term of the $\mathcal{N} = (1,0)$ Wess-Zumino-Witten model
\cite{WZW1, WZW2}. The other components of $v'$ do not appear in the final
action, so there is no need to include their kinetic terms. Note also that we have defined the kinetic 
term to have the
correct sign whether $k$ is positive or negative.
The Lagrangian for the boundary theory will now be given 
by a type of gauged
$\mathcal{N} = (1,0)$ WZW model
\begin{equation}
  \mathcal{L}_{b, k}(v', \Gamma') = \mathcal{L}_{kb, k}(v', \Gamma')
 + \mathcal{L}_{pb, k}(v', \Gamma'). \label{bt}
\end{equation}
and so the complete gauge and supersymmetry invariant action is given by
\begin{equation}
\mathcal{L}_{sgCS,  k}(v', \Gamma) = \mathcal{L}_{sCS,   k}(\Gamma) 
+ \mathcal{L}_{b, k}(v', \Gamma').
\end{equation}
The component form of $\mathcal{L}_{sCS,   k} + \mathcal{L}_{pb, k}$ 
 is obtained by substituting  
\begin{eqnarray}
A_\mu \rightarrow i \mu (\mathcal{D}_\mu \mu^{-1}), &&
\chi_a \rightarrow \mu \chi_a \mu^{-1} -i \psi_a, \nonumber \\
E_a \rightarrow \mu E_a \mu^{-1}, &&
\end{eqnarray}
where we have defined the components of $v$ to be
\begin{equation}
\mu = v_| \;\; , \;\; \psi_a = (D_a v)_| \mu^{-1},
\end{equation}
in the original SUSY boundary action,
\begin{eqnarray}
 \mathcal{L}_{sCS, k} &=& 
\frac{k}{4\pi} \left[ \epsilon^{\mu \nu \rho}
 \left( A_\mu \partial_\nu  A_\rho 
 + \frac{2i}{3} A_\mu A_\nu  A_\rho  \right)   \right. 
\nonumber \\  && \,\,\,\,\,\,\,\,\,\, \left. + 
 E^a E_a
 + \mathcal{D}_\mu ( \chi^a(\gamma^\mu)_a^b E_b) 
+ \mathcal{D}_3 (\chi^a E_b)
\right. 
\nonumber \\  && \,\,\,\,\,\,\,\,\,\, \left.
+  \frac{i}{6}\mathcal{D}_3  \left( \chi^a\left[(\gamma^\mu A_{\mu})_{a}^b ,\chi_b\right]\right)
\right].
\end{eqnarray}
So we can write 
\begin{eqnarray}
 \mathcal{L}_{sCS, k} + \mathcal{L}_{pb, k} &=& 
\frac{k}{4\pi} \left[ - \epsilon^{\mu \nu \rho}
 (\mu^{-1} \mathcal{D}_\mu \mu ) \partial_\nu (\mu^{-1} 
\mathcal{D}_\rho \mu )  
\right. 
\nonumber \\  && \,\,\,\,\,\,\,\,\,\, \left. 
 + \frac{2}{3}\epsilon^{\mu \nu \rho}(\mu^{-1} \mathcal{D}_\mu \mu )
(\mu^{-1} \mathcal{D}_\nu \mu )(\mu^{-1} \mathcal{D}_\rho \mu )
\right. 
\nonumber \\  && \,\,\,\,\,\,\,\,\,\, \left.  
-i \mathcal{D}_\mu ( \psi^a(\gamma^\mu)_a^b \mu E_b \mu^{-1}) 
-i \mathcal{D}_3 (\psi^a \mu E_b \mu^{-1}) \right. 
\nonumber \\  && \,\,\,\,\,\,\,\,\,\, \left.
+ \mathcal{D}_3 (\chi^a E_b) +  \frac{i}{6}\mathcal{D}_3  \left( (\mu\chi^a \mu^{-1} - i \psi^a)\right. \right.
\nonumber \\  && \,\,\,\,\,\,\,\,\,\, \left. \left. \times  \left[(\mu\gamma^\mu  \mathcal{D}_{\mu} \mu^{-1} )_{a}^b,  
 ( \mu\chi_b \mu^{-1}  - i \psi_b )\right] \right)
 \right. 
\nonumber \\  && \,\,\,\,\,\,\,\,\,\, \left.   + 
 E^a E_a   + \mathcal{D}_\mu ( \chi^a(\gamma^\mu)_a^b E_b) 
\right].
\end{eqnarray}
The component form of $\mathcal{L}_{kb, k} $ 
is the kinetic term for the $\mathcal{N} = (1,0)$
gauged WZW model,
\begin{eqnarray}
\mathcal{L}_{kb, k} &=&- \frac{|k|}{2\pi}
 \left[  -{\mu'}^{ -1} \psi_- \mathcal{D}_{+} (\psi_- \mu')
+ ( {\mu'}^{-1} \mathcal{D}_- \mu' )
({\mu'}^{-1} D_+ \mu' ) \right],
 \end{eqnarray}
where $\psi_- = (D_{-}\hat{v})_| {\mu'}^{-1}$ is the single fermionic component of $\hat{v}$.

Thus, $\mathcal{N} =1$ Chern-Simons
 theory in the presence of a boundary can 
be made both gauge and supersymmetry
invariant by 
the addition of a suitable theory on the boundary such that its gauge and
supersymmetry variations exactly cancel those of 
the Chern-Simons theory.  
Our result generalises that of \cite{21} which gave
the boundary terms to restore supersymmetry but not gauge invariance for
Chern-Simons theory, in the case of an Abelian gauge group.
It may be remarked that it was already known that the bosonic
Chern-Simons theory suitably
coupled to a gauged Wess-Zumino-Witten theory on the boundary
 is  gauge invariant \cite{20}, and
we have now provided a superspace extension of that result, or equivalently a
fully gauge invariant extension of the manifestly supersymmetric Chern-Simons with
boundary theories considered in \cite{MemBdry}.

\section{ABJM Theory}
\label{ABJM}

In the previous section we analysed $\mathcal{N} =1$ Chern-Simons theory in presence of a boundary. In this 
section we shall use the results of the previous section to analyse the ABJM theory in the presence of 
a boundary. 
The ABJM theory in the presence of a boundary, in $\mathcal{N} =1$ superspace formalism can be 
formulated as a  supersymmetric gauge theory with the gauge group $U(N)_k \times U(N)_{-k}$
and the superfield Lagrangian  
\begin{equation}
 \mathcal{L}_{ABJM, k} = \mathcal{L}_{CS,  k} (\Gamma) +
 \mathcal{L}_{CS,  -k} (\tilde\Gamma)  +
 \mathcal{L}_{M, k},
\end{equation}
 where $\mathcal{L}_{CS,  k} $ and $\mathcal{L}_{CS,  -k}$
 are Chern-Simons 
theories as discussed in the previous section, and
the matter part of the Lagrangian $\mathcal{L}_{M, k}$ is given by 
\begin{equation}
 \mathcal{L}_{M, k}  = 
 \mathcal{L}_{kM} + 
 \mathcal{L}_{pM, k}, 
\end{equation}
where  $\mathcal{L}_{pM, k}$ is the potential term given by 
\begin{eqnarray}
 \mathcal{L}_{pM, k} &=& 
-\frac{2\pi}{k}\nabla^2 
[\epsilon_{IJ}\epsilon^{KL}X^I Y_K X^{J}  Y_{L } \nonumber \\ && \,\,\,\,\,\,\,\,\,\,+
\epsilon^{IJ}\epsilon_{KL}X^{\dagger}_IY^{K \dagger}
 X^{J\dagger}  Y^{\dagger}_{L }]_|, 
\end{eqnarray}
and $\mathcal{L}_{kM}$ is the kinetic term 
given by 
\begin{eqnarray}
 \mathcal{L}_{kM}
&=&-\frac{1}{4}  \nabla^2 [ \nabla^a X^I  \nabla_a X^{\dagger}_I +
 \nabla^a Y^I \nabla_a Y^{\dagger}_I]_|.
\end{eqnarray}
Here the super-covariant derivatives for the matter fields are given by 
\begin{eqnarray}
 \nabla_a X^I &=&  D_a X^I + i\Gamma_a X^I - i X^I \tilde \Gamma_a, \nonumber \\ 
 \nabla_a Y^{I\dagger} &=&  D_a Y^{I\dagger} + i\Gamma_a Y^{I\dagger} 
- iY^{I\dagger} \tilde \Gamma_a ,\nonumber \\ 
 \nabla_a X^{I\dagger} &=&  D_a X^{I\dagger}  - iX^{I\dagger}  \Gamma_a  
+ i \tilde\Gamma_aX^{I\dagger} , \nonumber \\ 
 \nabla_a Y^{I} &=&  D_a Y^{I} - iY^{I} \Gamma_a  + 
i \tilde \Gamma_a Y^{I}.
\end{eqnarray}
The full finite gauge 
transformation under which the ABJM theory,
 without a boundary, is invariant are given by
\begin{eqnarray}
   \Gamma_a \rightarrow i u \, \nabla_a u^{-1},&&
   \tilde \Gamma_a \rightarrow i \tilde u \, \nabla_a \tilde u^{-1},\nonumber \\
 X^I \rightarrow  u  X^I \tilde u^{-1},&&
 X^{I\dagger} \rightarrow  \tilde u  X^{I\dagger}  u^{-1},\nonumber \\
 Y^I \rightarrow  \tilde u  Y^I  u^{-1},&&
 Y^I \rightarrow  u  Y^{I\dagger} \tilde u^{-1},
\end{eqnarray}
where 
\begin{eqnarray}
 u &=& \exp ( i \Lambda^A T_A), \nonumber \\
\tilde u &=& \exp ( i \tilde \Lambda^A T_A).
\end{eqnarray}
The infinitesimal gauge transformations of these fields are given by  
\begin{eqnarray}
 \delta \Gamma_a =  \nabla_a  \Lambda, 
&&   \delta \tilde\Gamma_a = \tilde\nabla_a 
 \tilde\Lambda, \nonumber \\ 
\delta X^{I } = i(\Lambda X^{I }  - X^{I }\tilde \Lambda ),  
&&  \delta  X^{I \dagger  } 
= i(   \tilde \Lambda X^{I\dagger  }-X^{I\dagger  }\Lambda) 
, \nonumber \\ 
\delta Y^{I } = i(   \tilde \Lambda Y^{I } -Y^{I }\Lambda ), 
 &&  \delta  Y^{I \dagger  } 
= i(\Lambda Y^{I\dagger  } - Y^{I\dagger  }\tilde \Lambda). \label{ghjkl}
\end{eqnarray}

We can now 
discuss the ABJM theory in the presence of  a boundary.
We can use the analysis in the previous section and a generalisation of the
work done in \cite{21, MemBdry, 20} to analyse the ABJM theory in the presence of a boundary.
The SUSY variation of the ABJM Lagrangian on a manifold with a boundary
is a boundary term. Thus, to retain the SUSY of the theory a suitable boundary piece 
has to be  added in such a way  that it cancels the boundary term generated by the 
 SUSY variation of the original theory. The sum of this boundary term and the original ABJM 
Lagrangian can now be written as  
\begin{equation}
 \mathcal{L}_{sABJM, k} = \mathcal{L}_{sCS,   k} (\Gamma) + 
 \mathcal{L}_{sCS,   -k} (\tilde\Gamma)  +
 \mathcal{L}_{sM, k},
\end{equation}
 where $\mathcal{L}_{sCS,  k} $ and $\mathcal{L}_{sCS, - k}$
 are the Chern-Simons theories
on a manifold with a boundary defined in the previous section.
The matter part of the Lagrangian $\mathcal{L}_{sM, k}$ is given by 
\begin{equation}
 \mathcal{L}_{sM, k} = 
 \mathcal{L}_{skM} +
 \mathcal{L}_{spM, k}, 
\end{equation}
where  $\mathcal{L}_{spM, k}$ is the potential term given by 
\begin{eqnarray}
 \mathcal{L}_{spM, k}&=& 
\frac{2\pi}{k}(-\nabla^2 + \mathcal{D}_3)
[\epsilon_{IJ}\epsilon^{KL}X^IY_K X^{J}  Y_{L } \nonumber \\ && +
\epsilon^{IJ}\epsilon_{KL}X^{\dagger}_IY^{K \dagger}X^{J\dagger} Y^{\dagger}_{L }]_|, 
\end{eqnarray}
and the kinetic term 
 $\mathcal{L}_{skM}$ is now given by   
\begin{eqnarray}
 \mathcal{L}_{skM}
&=& \frac{1}{4}  (-\nabla^2 + \mathcal{D}_3) [  \nabla^a X^I 
 \nabla_a X^{\dagger}_I
\nonumber \\ && +
   \nabla^a Y^I  \nabla_a Y^{\dagger}_I]_|.
\end{eqnarray}
The matter part of the ABJM theory is still invariant under the 
 gauge transformations given by Eq. $(\ref{ghjkl})$. 
However, the Chern-Simons part
 is not invariant under these 
gauge transformations. Thus, the  total 
 Lagrangian  for the ABJM theory 
is not invariant under the gauge 
transformations given by Eq. $(\ref{ghjkl})$.
However, this is exactly the issue we tackled 
in the previous section, so we
know that we can add a boundary 
action to modify the ABJM action. The result is
the supersymmetric and gauge invariant action:
\begin{equation}
\mathcal{L}_{sgABJM,  k} = \mathcal{L}_{sgCS,  k} (v', \Gamma) +
\mathcal{L}_{sgCS,  -k} (\tilde{v}', \tilde\Gamma)  +
 \mathcal{L}_{sM, k}.
\end{equation}
Furthermore, $v'$ and $\tilde v'$ can be extended in to
 the bulk to produce
fields $v$ and $\tilde v$ whose finite gauge transformations are given by
 \begin{eqnarray}
 v &\rightarrow& vu^{-1}, \nonumber \\
\tilde v &\rightarrow& \tilde v \tilde u^{-1}.
 \end{eqnarray}

Thus, by introducing new degrees of freedom on the boundary, we have found
a superspace description of the boundary ABJM theory which is also gauge
invariant.
It would be interesting to generalise this to extended superspace
\footnote{The supersymmetric but not gauge invariant case for $\mathcal{N}=2$ supersymmetry can be found in \cite{MemBdry}.}
 so that more
supersymmetry was manifest, and to investigate
 in detail how much supersymmetry
is preserved by this theory or similar supersymmetric
 Chern-Simons theories with
matter in the presence of a boundary.
Other than some technical complications, it should be possible to extend this
analysis to $\mathcal{N}=2$ superspace, and indeed
when the $\mathcal{N}=2$ Chern-Simons action was derived, its similarity to the
$\mathcal{N}=2$ WZW action was noted \cite{IvanovN2CS}. However, an
interesting question is whether the full supersymmetry will give further
constraints on the boundary action, as we seemingly have the freedom to add
any additional supersymmetric gauge invariant boundary terms. One
obvious question is whether the boundary theory relates the two $SU(N)$ factors
such as through a coupling which preserves the diagonal subgroup. Some such
feature may be expected as the $\mathcal{N}=6$ bulk ABJM action required
the specific $SU(N) \times SU(N)$ form of the gauge group, but this is not
required by less supersymmetric theories. Going beyond manifest $\mathcal{N}=2$
supersymmetry is even more difficult, but the ABJM action has been formulated
in $\mathcal{N}=3$ harmonic superspace \cite{1}. Alternatively it may
be possible to proceed without an off-shell superspace
action using the ectoplasm formalism \cite{Ecto1, Ecto2}, as recently explored
for systems with a boundary \cite{EctoEdge}.

\section{BRST and Anti-BRST Symmetries }
In this section we will study the BRST and anti-BRST symmetries
 of the theory discussed in the previous 
section. 
As the sum of the boundary theory and  ABJM theory is invariant 
under gauge transformations, 
it contains unphysical degrees of freedom. 
These unphysical degrees of freedom will give rise 
to constraints in canonical quantisation 
and divergences in the partition function in 
the path integral quantisation. So before we can 
quantise this theory we will need to 
eliminate these unphysical degrees of freedom by 
the addition of a suitable
gauge fixing term and a suitable ghost term to  it. 
The new effective Lagrangian which is 
obtained by taking the sum of the original classical Lagrangian, 
the gauge fixing term and the ghost term will  be invariant 
under two new sets of
 transformations called the BRST transformation and the 
anti-BRST transformation.

In order to write a suitable
gauge fixing term and a suitable ghost for the ABJM theory, we denote 
the  auxiliary  
superfields  by  $B, \tilde B$. We also denote the ghosts by 
$C, \tilde C$ and the anti-ghosts by $\overline{C}, 
\tilde{\overline{C}}$. It may be noted that whereas the auxiliary 
fields are regular 
 matrix  valued scalar superfields, the ghosts and the anti-ghosts 
 are matrix
 valued anti-commuting superfields. All these superfields are 
 suitably contracted with generators of the Lie algebra in the adjoint 
representation 
\begin{eqnarray}
 B = B^A T_A, && \tilde B = \tilde B^A T_A, \nonumber \\
\tilde C =\tilde C^A  T_A, && C =C^A  T_A, \nonumber \\ 
\overline{C}= \overline{C}^A T_A.&& \tilde{\overline{C}}=\tilde{\overline{C}}^A T_A. 
\end{eqnarray}  
Now we can write the 
 gauge-fixing term  $\mathcal{L}_{gf}$ and the ghost term 
  $\mathcal{L}_{gh}$ for 
the  ABJM theory corresponding to the 
 gauge-fixing function \cite{mf012}, 
\begin{eqnarray}
 D^a \Gamma_a=0, && D^a \tilde \Gamma_a =0,
\end{eqnarray}
as follows
\begin{eqnarray}
 \mathcal{L}_{gf} &=&- \nabla_+ \nabla_- [ B D^a \Gamma_a]_| + - \tilde \nabla_+ \tilde\nabla_- [ \tilde B
 D^a \tilde\Gamma_a]_| ,  \label{gt}\\ 
 \mathcal{L}_{gh}&=&- \nabla_+ \nabla_-[\overline{C} 
 D^a \nabla_a C ]_| +
\tilde \nabla_+ \tilde \nabla_-[\tilde{\overline{C}} 
 D^a \tilde \nabla_a \tilde C ]_|. \label{hgt}
\end{eqnarray}  
We now define an
  effective Lagrangian $\mathcal{L}_{eff, k}$  as  the sum
of the  supersymmetric and gauge invariant ABJM Lagrangian, 
the  gauge-fixing term  and the ghost term, 
\begin{equation}
 \mathcal{L}_{eff, k} =  
\mathcal{L}_{sgABJM,  k} + \mathcal{L}_{gf} +  \mathcal{L}_{gh}.
\end{equation}
The BRST transformations of the matter fields can be written as 
\begin{eqnarray}
s\, X^{I } = i(C X^{I } - X^{I } \tilde C ),  &&  s\,  X^{I \dagger  } 
= i(  \tilde C X^{I\dagger  } -X^{I\dagger  } C) 
, \nonumber \\ 
s\, Y^{I } = i(   \tilde C Y^{I }-Y^{I } C),  &&  s\,   Y^{I \dagger  } 
= i(C Y^{I\dagger  } - Y^{I\dagger  } \tilde C).
\end{eqnarray}
 The  BRST transformations of the auxiliary  
superfields, ghosts and anti-ghosts  
can be written as  
\begin{eqnarray}
s \,\Gamma_{a} = \nabla_a  C, && s \,\tilde\Gamma_{a} = \tilde\nabla_a  \tilde C, \nonumber \\ 
s \,C = - \frac{1}{2}\{C, C\} , && s \, \tilde C = - \frac{1}{2}
\{\tilde C , \tilde C\} \nonumber \\
s \,\overline{C}= B, && s \,\tilde{\overline{C}}= \tilde{B}, \nonumber \\ 
s \,B =0, && s\,\tilde B =0.
\end{eqnarray}
The BRST transformation of the $v$ and $ \tilde v$ can be written as 
\begin{eqnarray}
 s\, v = -i v C, &&s\,  \tilde v = -i \tilde v \tilde C. 
\end{eqnarray}
These BRST transformations are nilpotent and thus  satisfy 
$s^2 =0$.
This fact  can be used to show that the  
sum of the  gauge fixing term $\mathcal{L}_{gf} $
 and  the ghost term $ \mathcal{L}_{gh}$ 
is invariant under BRST transformations. 
It is because the sum of the ghost term and 
the gauge fixing term can be written as 
\begin{equation}
 \mathcal{L}_{gf} +  \mathcal{L}_{gh} = -  \nabla_+ \nabla_- s [ \overline{C} D^a \Gamma_a]_| 
+  \tilde\nabla_+ \tilde\nabla_- 
s [ \tilde{\overline{C}} D^a \tilde\Gamma_a]_|. 
\end{equation}
Now using  the fact that 
 BRST transformations are nilpotent, we get 
\begin{equation}
s \mathcal{L}_{gf} + s \mathcal{L}_{gh} 
= -  \nabla_+ \nabla_- s^2 [ \overline{C} D^a \Gamma_a]_|
+ \tilde \nabla_+ \tilde\nabla_- s^2 [ \tilde{\overline{C}}
 D^a \tilde\Gamma_a]_|=0. 
\end{equation}
The Lagrangian $\mathcal{L}_{sABJM, k}$
 is not invariant under these BRST transformations as it generates a 
boundary term  which is given by 
\begin{equation}
s\mathcal{L}_{sABJM, k} =  \frac{k}{2\pi} (P_- {\nabla}')^a  [ C' \omega_a'  ]_| -
\frac{k}{2\pi} (P_- {\tilde\nabla}')^a  [ \tilde C' \tilde \omega_a'  ]_|.
\end{equation}
Here $C' $ and $ \tilde C'$ are the induced values of 
$C $ and $ \tilde C$ on the
 boundary.  However, this boundary term is exactly cancelled by 
the BRST variation of boundary theory. Thus, the sum of the bulk and 
the boundary theory is invariant under these BRST transformations,
and so we have $  s \mathcal{L}_{sgABJM,  k} =0$. Thus, the
effective Lagrangian $\mathcal{L}_{eff, k}$ 
 is invariant under BRST transformations, 
\begin{equation}
s \mathcal{L}_{eff, k} =  
s\mathcal{L}_{sgABJM,  k} + s\mathcal{L}_{gf} +  s\mathcal{L}_{gh} =0.
\end{equation}

We can perform a similar analysis using the anti-BRST transformations. 
The anti-BRST transformations of the matter fields can be written as 
\begin{eqnarray}
\overline{s} \,X^{I } = i(\overline C X^{I } - X^{I } \tilde{\overline{C}}),
  &&  \overline{s} \,  X^{I \dagger  } 
= i( \tilde{\overline{C}}X^{I\dagger  } - X^{I\dagger  }\overline C  ), \nonumber \\ 
\overline{s} \, Y^{I } = i(   \tilde{\overline{C}}Y^{I }-Y^{I }\overline C ),  && \overline{s} \,  Y^{I \dagger  } 
= i(\overline C Y^{I\dagger  }-Y^{I\dagger  } \tilde{\overline{C}}). 
\end{eqnarray}
 The anti-BRST transformations of the auxiliary  
superfields, ghosts and anti-ghosts  can be 
written as 
\begin{eqnarray}
\overline{s} \,\Gamma_{a} = \nabla_a \overline{C}, &&\overline{s} \,\tilde \Gamma_{a} = \tilde
\nabla_a \tilde{\overline{C}}, \nonumber \\ 
\overline{s} \,C = -B - \{ \overline{C}, C\}, &&  \overline{s} \,\tilde C =
 -\tilde B -   \{ \tilde{\overline{C}} , \tilde C\},\nonumber \\
\overline{s} \,\overline{C} = - \frac{1}{2}\{\overline{C}, \overline{C}\},
&&\overline{s} \,\tilde{\overline{C}} = - 
\frac{1}{2}\{\tilde{\overline{C}}, \tilde{\overline{C}}\},
\nonumber \\
\overline{s} \,B = \frac{1}{2}[B , \overline{C}] && \overline{s} \,\tilde B =
 \frac{1}{2}\tilde [B, \tilde{\overline{C}}].
\end{eqnarray}
The BRST transformation of the $v$ and $\tilde v$ 
fields can be written as 
\begin{eqnarray}
 \overline{s} \, v = -i v \overline C, &&\overline{s}\,  \tilde v = -i \tilde v \tilde{\overline{C}}. 
\end{eqnarray}
The anti-BRST transformations also are nilpotent 
and thus  satisfy  $\overline{s}^2 =0$.  Furthermore, the 
sum of the ghost and gauge fixing terms 
 can also be written as  
\begin{equation}
   \mathcal{L}_{gf} +  \mathcal{L}_{gh} = 
-  \nabla_+ \nabla_-   \overline{s} [ \overline{C} D^a \Gamma_a]_|
+  \tilde\nabla_+ \tilde\nabla_-  
 \overline{s} [ \tilde{\overline{C}} D^a \tilde\Gamma_a]_|.
\end{equation}
Thus, using  the fact that 
 anti-BRST transformations are nilpotent, we get 
\begin{equation}
\overline{s} \mathcal{L}_{gf} + \overline{s} \mathcal{L}_{gh} 
= -  \nabla_+ \nabla_- \overline{s}^2 [ \overline{C} D^a \Gamma_a]_|
+ \tilde \nabla_+ \tilde\nabla_- \overline{s}^2 [ \tilde{\overline{C}}
 D^a \tilde\Gamma_a]_|=0. 
\end{equation}
Here again the Lagrangian $\mathcal{L}_{sABJM, k}$
 is not invariant under these anti-BRST transformations and it generates a 
boundary term  which is given by 
\begin{eqnarray}
  \overline{s} \mathcal{L}_{sABJM, k } &=& 
 \frac{k}{2\pi}(P_- {\nabla}')^a  [ \overline{C}' \omega_a' ]_|
- \frac{k}{2\pi}(P_- {\tilde\nabla}')^a 
 [ \tilde{\overline{C}}' \tilde\omega_a' ]_|.
\end{eqnarray}
Here $\overline C ' $ and $ \tilde{ \overline{C}}'$ are the induced values of 
$\overline C $ and $ \tilde{ \overline{C}}$ on the
 boundary.
This term is again cancelled by the anti-BRST variation of the boundary 
theory and so we have $  \overline{s} \mathcal{L}_{sgABJM,  k} =0$. 
Thus, the
effective Lagrangian $\mathcal{L}_{eff, k}$ 
 is also invariant under these anti-BRST transformations, 
\begin{equation}
\overline{s} \mathcal{L}_{eff, k} =  
\overline{s}\mathcal{L}_{sgABJM,  k} + 
\overline{s}\mathcal{L}_{gf} +  \overline{s}\mathcal{L}_{gh} =0.
\end{equation}
So the effective Lagrangian for the supersymmetric and gauge invariant ABJM
 is invariant under both the BRST and 
 the anti-BRST transformations. 
\section{Conclusion}
In this paper we analysed the $\mathcal{N}=1$ 
Chern-Simons theory in the presence of a boundary. We  used 
the results thus obtained  to study the
 ABJM theory  in the presence of a boundary.
We first modified the Chern-Simons theory by adding a   boundary 
term to it such that  supersymmetry variations of the bulk 
Chern-Simons theory were cancelled by the supersymmetry variations of
 this  boundary term. 
The resultant theory was then
made gauge invariant by
 adding new boundary degrees of freedom to it.
 This new boundary theory 
was identified as a gauged Wess-Zumino-Witten model. 
These results were used to obtain a 
 superspace description of the boundary ABJM theory which was also gauge
invariant.
As the matter part of the ABJM theory is gauge invariant even with a
boundary, it was only necessary to include a boundary term to restore SUSY.
The Chern-Simons part of the ABJM was modified by both the addition of
a term to make it SUSY 
and new boundary degrees of freedom to make it 
gauge invariant. 
Thus, we added  a suitable theory on the boundary such that its gauge and
supersymmetry variations exactly cancel those of 
the bulk ABJM theory. We also analysed the BRST and the anti-BRST 
symmetries of this resultant theory. 
 
Chern-Simons theories are also important in 
 condensed matter physics due to their relevance in
 to fractional quantum Hall effect
\cite{a,b,c,d}.   Fractional quantum Hall effects is based on the 
concept of statistical transmutation, i.e.
the fact that in two dimensions, fermions can be described as charged bosons carrying
an odd integer number of  flux quanta which is achieved by  analysing  
 Chern-Simons fields coupled to the bosons. In this theory electrons in an external
magnetic field are described as bosons in a combined external and statistical magnetic
field. At special values of the filling fraction the statistical field cancels the external field,
in the mean field sense and the system is described as a gas of bosons feeling no net
magnetic field. These bosons condense into a homogeneous ground state. 
This model describes the quantisation of the Hall conductance and the existence of vortex and anti-vortex excitations. 
Lately supersymmetric generalisation of fractional quantum Hall effect have also
 been investigated \cite{a1, a2, a3, a4}.
In particular physical properties of the topological
excitations in the supersymmetric quantum Hall liquid have been  discussed
 in a dual supersymmetric
Chern-Simons theory \cite{a5}. 
Boundary effects for Chern-Simons theories are also
important in condensed matter physics. This is because 
in quantum Hall systems  gapless edge
modes exist \cite{qh6}. These have  important consequences for the transport properties of
 the system \cite{qh7}.
 These modes have been studied in the presence of an infinitely steep
external confining potential  \cite{ qh9, qh10}. 
 The description of these modes  has also been related
 to the chiral Luttinger liquid description of the edge excitations~\cite{qh12}.  
Thus,  the  results of this paper will be 
useful in analysing the supersymmetric generalisation of 
gapless edge modes of fractional quantum Hall systems.
This can have  important consequences for the transport properties
 of these fractional quantum hall  system.

\section*{Acknowledgements}
DJS is supported in part by the STFC Consolidated Grant ST/J000426/1.

\appendix

\section{Component BRST Transformations}
In this appendix we will first study the gauge transformations of the ABJM theory and the boundary theory 
 in the component form. 
We will then  analyse the BRST and anti-BRST transformations of 
these theories in the component form.
To do so we  write ghosts, anti-ghosts and the auxiliary 
fields in component form as 
 \begin{eqnarray}
c = [C]_|, &\overline{c} = [\overline{C} ]_|,&b = [B]_|, \nonumber \\ 
 {c}_a = [\nabla_a C]_|, &\overline{c}_a = [\nabla_a \overline{C}]_|, &  b_a = [\nabla_a B]_|,
 \nonumber \\ 
 {\rm{c}} =  [ \nabla^2 C  ]_|, & 
\overline{\rm{c}} =  [ \nabla^2 \overline{C} ]_|,& 
 {\rm{b}} =  [ \nabla^2  B  ]_|,\nonumber \\ 
\tilde{c} = [\tilde C]_|, &\tilde{\overline{c}} = [\tilde{\overline{C}} ]_|,&\tilde b = [\tilde B]_|, \nonumber \\ 
 \tilde c_a = [\tilde \nabla_a \tilde C]_|, &\tilde{\overline{c}}_a = [\tilde\nabla_a \tilde{\overline{C}}
]_|, &  \tilde b_a = [\tilde \nabla_a \tilde B]_|, \nonumber \\ 
 \tilde{\rm{c}} =  [ \tilde \nabla^2 \tilde C  ]_|, &
\tilde{\overline{{\rm{c}}}} =  [ \tilde\nabla^2 \tilde{\overline{C}} ]_|,& 
 \tilde{{\rm{b}}} =  [ \tilde\nabla^2  \tilde{B}  ]_|,\label{comp}
\end{eqnarray}
where the fields $c, \overline{c},{\rm{c}}, \overline{\rm{c}}, b_a$ and $
\tilde c, \tilde{\overline{c}},\tilde{\rm{c}}, \tilde{\overline{\rm{c}}}, \tilde b_a$ 
are fermionic fields and the fields $c_a, \overline{c}_a, b,  {\rm{b}} $ and $ 
\tilde c_a, \tilde{\overline{c}}_a, b,  \tilde{\rm{b}} $ 
are bosonic fields.
The components of the matter fields are given by  
 \begin{eqnarray}
 x^I = [X^I]_|, & x_a^I = [ \nabla_a X^I]_|, & {\rm{x}}^I = [{\nabla^2 {X}}^I]_|, \nonumber\\ 
 y^I = [Y^I]_|, & y_a^I = [ \nabla_a Y^I]_|, & {\rm{y}}^I = [{ \nabla^2 {Y}}^I]_|, \nonumber\\ 
 x   ^{I \dagger} = [X   ^{I \dagger}]_|, & x_a   ^{I \dagger} =
 [ \nabla_a X   ^{I \dagger}]_|, & {\rm{x}}   ^{I \dagger} = [{ \nabla^2 {X}}   ^{I \dagger}]_|, \nonumber\\ 
 y   ^{I \dagger} =
 [Y   ^{I \dagger}]_|, & y_a   ^{I \dagger} = [ \nabla_a Y   ^{I \dagger}]_|, & {\rm{y}}   ^{I \dagger} 
= [{ \nabla^2 {Y}}   ^{I \dagger}]_|. 
\end{eqnarray}
We also write the components of $\Lambda$ and $\tilde \Lambda $ as 
\begin{eqnarray}
{\lambda} = [\Lambda]_|, && \lambda_a = [\nabla_a\Lambda]_|, \nonumber \\ 
\overline{\lambda} = [\nabla^2 \Lambda]_|, &&  \tilde{\overline{\lambda}} = [\tilde{\nabla}^2 \tilde \Lambda]_|, \nonumber \\ 
 \tilde \lambda = [\tilde \Lambda]_|, && \tilde \lambda_a = [\tilde \nabla_a\tilde \Lambda]_|.
\end{eqnarray}
The component form of $v$ and $\tilde v$ are given by 
\begin{eqnarray}
\mu  = [v ]_|, && \mu_a  = [\nabla_a v]_|, \nonumber \\ 
 \nu  = [\nabla^2 v ]_|, && \tilde \nu  = [\tilde \nabla^2
 \tilde v]_|, \nonumber \\ 
\tilde \mu  = [\tilde v ]_|, && \tilde\mu_a  = [\tilde\nabla_a\tilde
v]_|.
\end{eqnarray}
The component form of $\Gamma_a$ and $ \tilde \Gamma_a $ are given by 
\begin{eqnarray}
 \chi_a = [\Gamma_a]_|, && A = - \frac{1}{2}[\nabla^a \Gamma_a]_|, \nonumber \\ 
A^\mu = - \frac{1 }{2} [ \nabla^a (\gamma^{\mu })_a^b \Gamma_b ]_|, &&
 E_a = - [\nabla^b \nabla_a  \Gamma_b]_|, \nonumber \\ 
\tilde \chi_a = [\tilde \Gamma_a]_|, && \tilde A = - \frac{1}{2}[\tilde \nabla^a \tilde \Gamma_a]_|, 
\nonumber \\ 
\tilde A^\mu = - \frac{1 }{2} [ \tilde \nabla^a (\gamma^{\mu })_a^b \tilde \Gamma_b ]_|, &&
 \tilde E_a = - [\tilde \nabla^b \tilde \nabla_a  \tilde \Gamma_b]_|.
\end{eqnarray}

Now after writing the components for all superfields we can 
write the gauge transformations of these component fields. 
Thus,  the component form of the gauge transformations
 of matter fields for the  ABJM theory are given by
\begin{eqnarray}
   {\delta }   \, x^{I }  &=& i(   {\lambda}x^{I } -
 x^{I } \tilde{   {\lambda}} ),  
\nonumber \\     
 {\delta }   \,  x^{I \dagger  } 
&=& -i(x^{I\dagger  } \lambda - \tilde{   {\lambda}}x^{I\dagger  }), 
\nonumber \\ 
   {\delta }   \, y^{I } &=& -i(y^{I }\lambda  - 
\tilde{   {\lambda}} y^{I }),  
\nonumber \\     
 {\delta }   \,   y^{I \dagger  } 
&=& i(\lambda y^{I\dagger  } - y^{I\dagger  } \tilde{   {\lambda}}), 
\nonumber \\ 
   {\delta }   \, x_a^{I } &=& i(\lambda_a x^{I } - x^{I }
\tilde{   {\lambda}}_a ) - i(\lambda x_a^{I } 
 - x_a^{I } \tilde{   {\lambda}} ), 
 \nonumber \\ 
     {\delta }   \,  x_a^{I \dagger  } 
&=& -i(x^{I\dagger  }\lambda_a - \tilde{   {\lambda}}_a x^{I\dagger  }) 
 +i(x_a^{I\dagger  } \lambda - \tilde{   {\lambda}}x_a^{I\dagger  }), 
\nonumber \\
   {\delta }   \, y_a^{I } &=&
 -i(y^{I }\lambda_a  -  \tilde{   {\lambda}}_a y^{I })
 +i(y_a^{I } \lambda  - \tilde{   {\lambda}} y_a^{I } ) , 
 \nonumber \\   
  {\delta }   \,   y_a^{I \dagger  } 
&=& i(\lambda_a y^{I\dagger  } - y^{I\dagger  } \tilde{   {\lambda}}_a) 
 - i(\lambda y_a^{I\dagger  } - y_a^{I\dagger  } \tilde{   {\lambda}}), 
\nonumber\\ 
   {\delta }   \, {\rm{x}}^{I } &=& i({\overline{\lambda}} x^{I } - x^{I } 
\tilde {\overline{\lambda}} ) + i(\lambda {\rm{x}}^{I } - {\rm{x}}^{I }  \tilde{  
{\lambda}} )
 - 2i (\lambda^a x_a^{I } - x^{a I } \tilde{   {\lambda}}_a ), 
 \nonumber \\  
   {\delta }   \,  {\rm{x}}^{I \dagger  } 
&=& -i(x^{I\dagger  } {\overline{\lambda}} - \tilde{\overline{\lambda}}x^{I\dagger  }) 
-i({\rm{x}}^{I\dagger  }\lambda - \tilde{   {\lambda}}{\rm{x}}^{I\dagger  }) 
 + 2i(x^{a I\dagger  } \lambda_a - \tilde{   {\lambda}}^a x_a^{I\dagger  }),
 \nonumber \\ 
   {\delta }   \, {\rm{y}}^{I } &=& 
-i(y^{I }{\overline{\lambda}}  - \tilde{\overline{\lambda}} y^{I })
-i({\rm{y}}^{I } \lambda  - \tilde{   {\lambda}}{\rm{y}}^{I } )
+2i(y^{a I } \lambda_a  - \tilde{   {\lambda}}^a y_a^{I }), 
\nonumber \\  
    {\delta }   \,   {\rm{y}}^{I \dagger  } 
&=& i({\overline{\lambda}}y^{I\dagger  } - y^{I\dagger  }\tilde{\overline{\lambda}}) 
 +i(\lambda {\rm{y}}^{I\dagger  } - {\rm{y}}^{I\dagger  }\tilde{   {\lambda}}) 
 -2 i(\lambda^a y_a^{I\dagger  }- y^{a I\dagger  }\tilde{   {\lambda}}_a).
\end{eqnarray}
The component form of the gauge transformation of 
the gauge fields for the  ABJM theory are given by 
\begin{eqnarray}
 \delta  \, \chi_a =  \chi_a\lambda +  \lambda_a , &&\delta  \,A = A \lambda +  {\overline{\lambda}}, \nonumber \\ 
\delta  \,A^\mu = \mathcal{D}_\mu \lambda , && \delta  \, E_a = E_a\lambda , \nonumber \\
 \delta  \, \tilde \chi_a = \tilde \chi_a\tilde  \lambda + \tilde \lambda_a , &&
\delta  \,\tilde A =\tilde A \tilde \lambda + \tilde{\overline{\lambda}}, \nonumber \\ 
\delta  \,\tilde A^\mu = \tilde{\mathcal{D}}_\mu \tilde \lambda , 
&& \delta  \, \tilde E_a = \tilde E_a \tilde  \lambda. 
\end{eqnarray}
The component form of the  gauge transformations
 for $v $ and $\tilde v $ are given by 
\begin{eqnarray} 
\delta  \, \mu = -i  \mu \lambda, 
&&
\delta  \, {\nu} = -i\nu \lambda -2i {\mu^a} \lambda_a  
-i  \mu {\overline{\lambda}},
\nonumber \\ 
\delta  \, \tilde \mu = -i  \tilde \mu \tilde \lambda,
&&
\delta  \, \tilde{\nu} = -i\tilde\nu \tilde\lambda
 -{2i {\tilde{\mu}^a}} \tilde\lambda_a  
-i  \tilde\mu \tilde{{\overline{\lambda}}},
\nonumber \\
 \delta \,  \mu_a = -i 
 \mu_a \lambda -i 
\mu \lambda_a,
&& 
\delta \,  \tilde\mu_a = -i 
 \tilde \mu_a \tilde \lambda -i 
\tilde \mu \tilde \lambda_a.
\end{eqnarray}

After discussing the component form of the gauge transformations, 
we will analyse the component form 
of the BRST and the anti-BRST transformations. 
In component form the BRST transformation of 
the matter fields in the  ABJM theory are given by 
\begin{eqnarray}
   {s }   \, x^{I }  &=& i(   {c}x^{I } -
 x^{I } \tilde{   {c}} ),  
\nonumber \\     
 {s }   \,  x^{I \dagger  } 
&=& -i(x^{I\dagger  } c - \tilde{   {c}}x^{I\dagger  }), 
\nonumber \\ 
   {s }   \, y^{I } &=& -i(y^{I }c  - 
\tilde{   {c}} y^{I }),  
\nonumber \\     
 {s }   \,   y^{I \dagger  } 
&=& i(c y^{I\dagger  } - y^{I\dagger  } \tilde{   {c}}), 
\nonumber \\ 
   {s }   \, x_a^{I } &=& i(c_a x^{I } - x^{I }
\tilde{   {c}}_a ) - i(c x_a^{I } 
 - x_a^{I } \tilde{   {c}} ), 
 \nonumber \\ 
     {s }   \,  x_a^{I \dagger  } 
&=& -i(x^{I\dagger  }c_a - \tilde{   {c}}_a x^{I\dagger  }) 
 +i(x_a^{I\dagger  } c - \tilde{   {c}}x_a^{I\dagger  }), 
\nonumber \\
   {s }   \, y_a^{I } &=&
 -i(y^{I }c_a  -  \tilde{   {c}}_a y^{I })
 +i(y_a^{I } c  - \tilde{   {c}} y_a^{I } ) , 
 \nonumber \\   
  {s }   \,   y_a^{I \dagger  } 
&=& i(c_a y^{I\dagger  } - y^{I\dagger  } \tilde{   {c}}_a) 
 - i(c y_a^{I\dagger  } - y_a^{I\dagger  } \tilde{   {c}}), 
\nonumber\\ 
   {s }   \, {\rm{x}}^{I } &=& i({\rm{c}} x^{I } - x^{I } 
\tilde {\rm{c}} ) + i(c {\rm{x}}^{I } - {\rm{x}}^{I }  \tilde{  
{c}} )
 - 2i (c^a x_a^{I } - x^{a I } \tilde{   {c}}_a ), 
 \nonumber \\  
   {s }   \,  {\rm{x}}^{I \dagger  } 
&=& -i(x^{I\dagger  } {\rm{c}} - \tilde{\rm{c}}x^{I\dagger  }) 
-i({\rm{x}}^{I\dagger  }c - \tilde{   {c}}{\rm{x}}^{I\dagger  }) 
 + 2i(x^{a I\dagger  } c_a - \tilde{   {c}}^a x_a^{I\dagger  }),
 \nonumber \\ 
   {s }   \, {\rm{y}}^{I } &=& 
-i(y^{I }{\rm{c}}  - \tilde{\rm{c}} y^{I })
-i({\rm{y}}^{I } c  - \tilde{   {c}}{\rm{y}}^{I } )
+2i(y^{a I } c_a  - \tilde{   {c}}^a y_a^{I }), 
\nonumber \\  
    {s }   \,   {\rm{y}}^{I \dagger  } 
&=& i({\rm{c}}y^{I\dagger  } - y^{I\dagger  }\tilde{\rm{c}}) 
 +i(c {\rm{y}}^{I\dagger  } - {\rm{y}}^{I\dagger  }\tilde{   {c}}) 
 -2 i(c^a y_a^{I\dagger  }- y^{a I\dagger  }\tilde{   {c}}_a).
\end{eqnarray}
The anti-BRST transformation of the matter fields in the ABJM theory in component form 
are given by 
 \begin{eqnarray}
  {\overline{s}}    \, x^{I }  &=& i(   {\overline{c}}x^{I } -
 x^{I } \tilde{   {\overline{c}}} ),  
\nonumber \\     
{\overline{s}}    \,  x^{I \dagger  } 
&=& -i(x^{I\dagger  } \overline{c} - \tilde{   {\overline{c}}}x^{I\dagger  }), 
\nonumber \\ 
  {\overline{s}}    \, y^{I } &=& -i(y^{I }\overline{c}  - 
\tilde{   {\overline{c}}} y^{I }),  
\nonumber \\     
{\overline{s}}    \,   y^{I \dagger  } 
&=& i(\overline{c} y^{I\dagger  } - y^{I\dagger  } \tilde{   {\overline{c}}}), 
\nonumber \\ 
  {\overline{s}}    \, x_a^{I } &=& i(\overline{c}_a x^{I } - x^{I }
\tilde{   {\overline{c}}}_a ) - i(\overline{c} x_a^{I } 
 - x_a^{I } \tilde{   {\overline{c}}} ), 
 \nonumber \\ 
    {\overline{s}}    \,  x_a^{I \dagger  } 
&=& -i(x^{I\dagger  }\overline{c}_a - \tilde{   {\overline{c}}}_a x^{I\dagger  }) 
 +i(x_a^{I\dagger  } \overline{c} - \tilde{   {\overline{c}}}x_a^{I\dagger  }), 
\nonumber \\
  {\overline{s}}    \, y_a^{I } &=&
 -i(y^{I }\overline{c}_a  -  \tilde{   {\overline{c}}}_a y^{I })
 +i(y_a^{I } \overline{c}  - \tilde{   {\overline{c}}} y_a^{I } ) , 
 \nonumber \\   
 {\overline{s}}    \,   y_a^{I \dagger  } 
&=& i(\overline{c}_a y^{I\dagger  } - y^{I\dagger  } \tilde{   {\overline{c}}}_a) 
 - i(\overline{c} y_a^{I\dagger  } - y_a^{I\dagger  } \tilde{   {\overline{c}}}), 
\nonumber\\ 
  {\overline{s}}    \, {\rm{x}}^{I } &=& i({\rm{\overline{c}}} x^{I } - x^{I } 
\tilde {\rm{\overline{c}}} ) + i(\overline{c} {\rm{x}}^{I } - {\rm{x}}^{I }  \tilde{  
{\overline{c}}} )
 - 2i (\overline{c}^a x_a^{I } - x^{a I } \tilde{   {\overline{c}}}_a ), 
 \nonumber \\  
  {\overline{s}}    \,  {\rm{x}}^{I \dagger  } 
&=& -i(x^{I\dagger  } {\rm{\overline{c}}} - \tilde{\rm{\overline{c}}}x^{I\dagger  }) 
-i({\rm{x}}^{I\dagger  }\overline{c} - \tilde{   {\overline{c}}}{\rm{x}}^{I\dagger  }) 
 + 2i(x^{a I\dagger  } \overline{c}_a - \tilde{   {\overline{c}}}^a x_a^{I\dagger  }),
 \nonumber \\ 
  {\overline{s}}    \, {\rm{y}}^{I } &=& 
-i(y^{I }{\rm{\overline{c}}}  - \tilde{\rm{\overline{c}}} y^{I })
-i({\rm{y}}^{I } \overline{c}  - \tilde{   {\overline{c}}}{\rm{y}}^{I } )
+2i(y^{a I } \overline{c}_a  - \tilde{   {\overline{c}}}^a y_a^{I }), 
\nonumber \\  
   {\overline{s}}    \,   {\rm{y}}^{I \dagger  } 
&=& i({\rm{\overline{c}}}y^{I\dagger  } - y^{I\dagger  }\tilde{\rm{\overline{c}}}) 
 +i(\overline{c} {\rm{y}}^{I\dagger  } - {\rm{y}}^{I\dagger  }\tilde{   {\overline{c}}}) 
 -2 i(\overline{c}^a y_a^{I\dagger  }- y^{a I\dagger  }
\tilde{   {\overline{c}}}_a). 
\end{eqnarray}
In component form the BRST transformations of gauge fields, 
ghosts, anti-ghosts and auxiliary fields for the ABJM theory are given by   
\begin{eqnarray}
 s \, \chi_a =  \chi_a c + c_a, &&s \,A =  Ac + {\rm{c}}, \nonumber \\ 
s \,A^\mu = \mathcal{D}_\mu c , && s \, E_a = E_a c, \nonumber \\
s \, c = - \frac{1}{2}\{c , c\},  && s \, c_a =  [ c , c_a], \nonumber \\ 
 s \, \overline{c}= b, &&  s \,  {\rm{c}} =  [c^a , c_a] - 
\{   c, {\rm{c}}\},\nonumber \\ 
s \, \overline{c}_a =  b_a,
&&  s \,  \rm{\overline{c}} = {\rm{b}}\nonumber \\ 
 s \, \tilde \chi_a = \tilde \chi_a \tilde c + \tilde c_a, &&s \,\tilde A = \tilde A\tilde  c +\tilde {\rm{c}}, \nonumber \\ 
s \,\tilde A^\mu = \tilde{\mathcal{D}}_\mu \tilde c , && s \, \tilde E_a = \tilde E_a \tilde c, \nonumber \\
s \, \tilde c = -\frac{1}{2} \{\tilde c , \tilde c\},  && s \, \tilde c_a =  
[\tilde c , \tilde c_a], \nonumber \\ 
 s \, \tilde{\overline{c}}= \tilde b, &&  s \,  \tilde {\rm{c}} =
 [ \tilde c^a , \tilde c_a] -    
\{\tilde c, \tilde{\rm{c}}\},\nonumber \\ 
s \, \tilde{\overline{c}}_a = \tilde b_a,
&&  s \,  \tilde{\rm{\overline{c}}} = \tilde{\rm{b}}\nonumber \\ 
 s \, b= 0, &&s \,b_a = 0, \nonumber \\ 
 s \, \tilde b= 0, &&s \,\tilde b_a = 0, \nonumber \\ 
s \,\tilde{b} = 0 && s \,\tilde{b} = 0.
\end{eqnarray}
The anti-BRST transformation of the gauge fields,  ghosts, anti-ghosts  and the auxiliary fields
 for the 
 ABJM theory 
in component form are given by 
\begin{eqnarray}
\overline{s}\, c = -b - \{ \overline{c}, c\},  && \overline{s}\, c_a = -b_a -[  \overline{c}_a , c]
 +  [ \overline{c}, c_a], \nonumber \\ 
 \overline{s}\, \overline{c}= -\frac{1}{2} \{\overline{c}, \overline{c}\},
 &&\overline{s} \,  
{\rm{c}} = - {\rm{b}} -  \{ {\rm{\overline{c}} } , c\}  - 
\{c, {\rm{\overline{c}}}\} + 2[\overline{c}^a, c_a],
  \nonumber \\ 
\overline{s}\, \overline{c}_a =  [ \overline{c}, \overline{c}_a],
&&  \overline{s} \,  {\rm{\overline{c}}} 
= [  \overline{c}^a, \overline{c}_a] -
 \{  \overline{c}, {\rm{\overline{c}}}\}, \nonumber \\ 
 \overline{s}\,b_a = \frac{1}{2}\{b_a, \overline{c}\} 
+ \frac{1}{2}[b, \rm{\overline{c}}_a],  &&
\overline{s} \,{\rm{b}} = \frac{1}{2}[{\rm{b}}, \overline{c}]  +\frac{1}{2}
[ b, {\rm{\overline{c}}}] + [ b^a, \overline{c}_a],\nonumber \\ 
 \overline{s} \, b= \frac{1}{2}[b, \overline{c}],  &&
\overline{s} \, \chi_a = \chi_a\overline{c} +  \overline{c}_a, \nonumber \\
\overline{s} \,A =  A\overline{c} +  {\rm{\overline{c}}}, && 
\overline{s} \,A^\mu = \mathcal{D}_\mu \overline{c},   \nonumber \\
\overline{s}\, \tilde c = -\tilde b - 
 \{\tilde{\overline{c}}, \tilde c\},  
&& \overline{s}\, \tilde c_a = -\tilde b_a - 
[\tilde{\overline{c}}_a, \tilde c]
 +  [ \tilde{\overline{c}}, \tilde c_a], \nonumber \\ 
 \overline{s}\, \tilde{\overline{c}}= - \frac{1}{2}
\{\tilde{\overline{c}}, \tilde{\overline{c}}\}, &&\overline{s} 
\,  
\tilde {\rm{c}} = - \tilde {\rm{b}} -  
\{\tilde{{\rm{\overline{c}} } },\tilde c\}  -   
 \{\tilde c, \tilde{{\rm{\overline{c}}}}\} +
 2 [\tilde{\overline{c}}^a, \tilde c_a],
  \nonumber \\ 
\overline{s}\, \tilde{\overline{c}}_a =  [ \tilde{\overline{c}},
 \tilde{\overline{c}}_a],
&&  \overline{s} \,  \tilde{{\rm{\overline{c}}}} = 
  [\tilde{\overline{c}}^a, \tilde{\overline{c}}_a] -
   \{\tilde{\overline{c}}, \tilde{{\rm{\overline{c}}}}\}, \nonumber \\ 
 \overline{s}\,\tilde b_a = \frac{1}{2}\{\tilde b_a, \tilde{\overline{c}}\}
 + \frac{1}{2}[\tilde b, \tilde{
\rm{\overline{c}}}_a],  &&
\overline{s} \,\tilde {\rm{b}} = \frac{1}{2}[
\tilde {\rm{b}}, \tilde{\overline{c}}]  + \frac{1}{2}[\tilde b,
\tilde{{\rm{\overline{c}}}}]
 +  [\tilde b^a, \tilde{\overline{c}}_a],\nonumber \\ 
 \overline{s} \, \tilde b= \frac{1}{2}[\tilde b, \tilde{\overline{c}}],  &&
\overline{s} \, \tilde \chi_a = \tilde \chi_a \tilde{\overline{c}} + \tilde{\overline{c}}_a, \nonumber \\
\overline{s} \,\tilde A = \tilde A \tilde{\overline{c}} + \tilde{{\rm{\overline{c}}}}, && 
\overline{s} \,\tilde A^\mu = \tilde{\mathcal{D}}_\mu \tilde{\overline{c}}, 
 \nonumber \\ \overline{s}\, \tilde{E}_a =   \tilde{E}_a \tilde{\overline{c}} , && \overline{s}\, E_a =  E_a \overline{c}.
\end{eqnarray}
Furthermore, the BRST transformations of $v , \tilde v $ 
in component form are given by 
\begin{eqnarray} 
s  \, \mu = -i  \mu c, 
&&
s  \, {\nu} = -i\nu c -2i {\mu^a} c_a  
-i  \mu {\rm{c}},
\nonumber \\ 
s  \, \tilde \mu = -i  \tilde \mu \tilde c,
&&
s  \, \tilde{\nu} = -i\tilde\nu \tilde c
 -{2i {\tilde{\mu}^a}} \tilde c_a  
-i  \tilde\mu \tilde{{\rm{c}}},
\nonumber \\
 s \, \mu_a = -i 
 \mu_a c -i 
\mu c_a,
&& 
s \,  \tilde\mu_a = -i 
 \tilde \mu_a \tilde c -i 
\tilde \mu \tilde c_a.
\end{eqnarray}
and the anti-BRST transformations of $v, \tilde v$ 
in component form are given by 

\begin{eqnarray} 
\overline{s}  \, \mu = -i  \mu { \overline{c}}, 
&&
\overline{s}  \, {\nu} = -i\nu { \overline{c}} -2i {\mu^a} { \overline{c}}_a  
-i  \mu {\rm{{ \overline{c}}}},
\nonumber \\ 
\overline{s}  \, \tilde \mu = -i  \tilde \mu \tilde { \overline{c}},
&&
\overline{s}  \, \tilde{\nu} = -i\tilde\nu \tilde{ \overline{c}}
 -{2i {\tilde{\mu}^a}} \tilde { \overline{c}}_a  
-i  \tilde\mu \tilde{{\rm{{ \overline{c}}}}},
\nonumber \\
 \overline{s} \,  \mu_a = -i 
 \mu_a { \overline{c}} -i 
\mu { \overline{c}}_a,
&& 
\overline{s} \,  \tilde\mu_a = -i 
 \tilde \mu_a \tilde { \overline{c}} -i 
\tilde \mu \tilde { \overline{c}}_a.
\end{eqnarray}
 These are  the component form of the  BRST and the
anti-BRST transformations of the ABJM theory and the 
boundary theory it is coupled to.

\end{document}